\documentclass[aps,prev,twocolumn,preprintnumbers,floatfix,nofootinbib]{revtex4-1}
\pdfoutput=1
\usepackage{graphicx}
\usepackage{bm}
\usepackage{times}
\usepackage{slashed}
\usepackage{color}
\usepackage{color}

\newcommand{\be}{\begin{equation}}
\newcommand{\ee}{\end{equation}}
\newcommand{\bea}{\begin{eqnarray}}
\newcommand{\eea}{\end{eqnarray}}
\newcommand{\mdm}{m_{H^0}}
\newcommand{\gev}{\mathrm{GeV}}

\begin{document}
\title{The CTA aims at the Inert Doublet Model}

\author{Farinaldo S. Queiroz}
\email{farinaldo.queiroz@mpi-hd.mpg.de}

\author{Carlos E. Yaguna}
\email{carlos.yaguna@mpi-hd.mpg.de}

\affiliation{Max-Planck-Institut fur Kernphysik, Saupfercheckweg 1, 69117 Heidelberg, Germany}

\begin{abstract}
We show that the Cherenkov Telescope Array (CTA) can realistically challenge the Inert Doublet Model, one of the simplest and best known models of dark matter. Specifically, the CTA may exclude its heavy regime up to  dark matter masses of 800 GeV and probe a large fraction of the remaining viable parameter space at even higher masses.   Two features of the Inert Doublet Model make it particularly suitable for CTA searches. First, the dark matter mass (in the heavy regime) must be larger than 500 GeV. Second, the dark matter annihilation cross section, $\sigma v$, is \emph{always} larger than the thermal one, reaching values as high as $10^{-25} \mathrm{cm^3s^{-1}}$. This higher value of  $\sigma v$ is the result of the unavoidable coannihilation effects that determine the relic density via thermal freeze-out in the early Universe.  We find that with 100 hours of Galactic Center exposure,  CTA's expected limit widely surpasses, even after the inclusion of systematic errors, current and projected bounds from Fermi-LAT and HESS on this model. 
\end{abstract}


\maketitle

\section{Introduction}
The indirect detection of dark matter is one of the most promising alternatives to observe and  identify the dark matter particle \cite{Buckley:2013bha}. Among the different channels that could give rise to an indirect detection signal, gamma rays \cite{Conrad:2011na} have the advantage of being easier to detect than neutrinos and of not being affected by propagation effects, unlike positrons and antiprotons. In fact, gamma-ray observations by Fermi-LAT currently provide the most stringent bounds on the dark matter annihilation rate \cite{Ackermann:2015zua}. They exclude a thermal cross section ($\sigma v=3\times 10^{-26}\mathrm{cm^3s^{-1}}$) up to dark matter masses of order 100 GeV. At TeV-scale masses, the strongest constraint comes instead from HESS \cite{Abramowski:2011hc}  --an Imaging Air Cherenkov Telescope (IACT)--, but it lies well above the thermal cross section. 

A major step forward in gamma-ray astrophysics will be the construction of the Cherenkov Telescope Array (CTA) \cite{Acharya:2013sxa,Persic:2013aoa}, which should start operating in 2019 and whose sensitivity is expected  to be significantly better than currently operating IACTs. Several works have already studied the CTA sensitivity to dark matter annihilations \cite{Bergstrom:2010gh,Bringmann:2011ye,Cirelli:2012ut,Ripken:2012db, Doro:2012xx,Wood:2013taa,Pierre:2014tra,Lefranc:2015pza,Silverwood:2014yza,Roszkowski:2014iqa,Carr:2015hta,Ibarra:2015tya}. In \cite{Silverwood:2014yza,Lefranc:2015pza} it was recently found  that systematic  errors  substantially degrade the CTA sensitivity and should, therefore, always be included in realistic assessments. At the same time, they found that  a spectral and morphological analysis substantially improves the CTA sensitivity to a dark matter signal from the Galactic center. When all these effects are taken into account, the picture that emerges is not overly optimistic. On the one hand, the CTA is expected to provide the strongest bounds on  the dark matter annihilation cross section at high masses. On the other hand, the expected limit, for a Einasto profile, lies above the thermal annihilation cross section for all annihilation channels but $\tau^+\tau^-$. Probing the thermal cross section for standard annihilation channels, such as $b\bar b$ or $W^+W^-$, would entail a DM profile steeper than NFW or Einasto. Thus, it would seem that a particularly favorable dark matter distribution in our Galaxy were required for the CTA to play an important role in dark matter searches.

\begin{figure}[!t]
\includegraphics[width=\columnwidth]{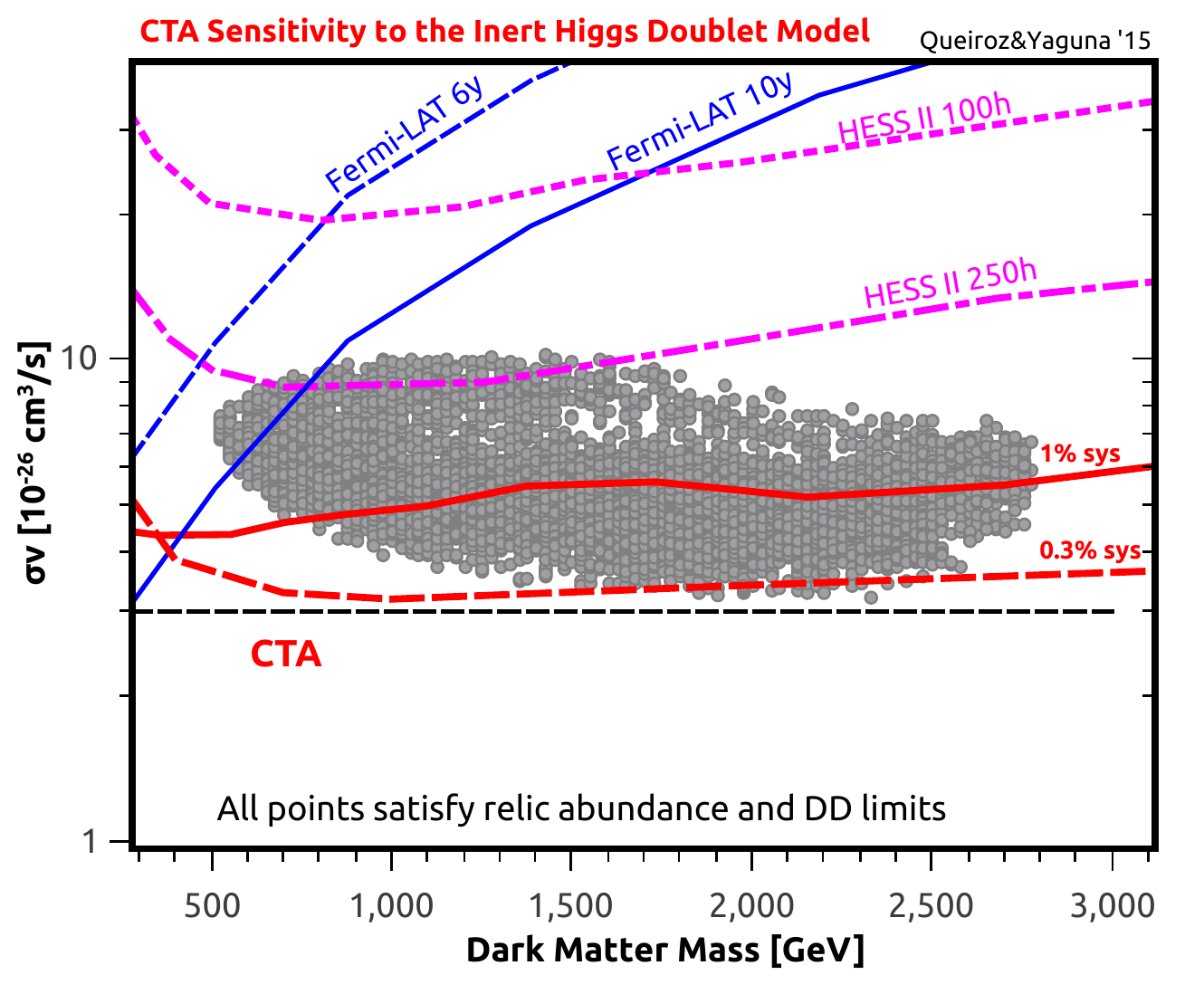}
\caption{The CTA sensitivity to the IDM. All points in the figure are consistent with the observed dark matter  abundance and with current LUX limits. It is clear from the figure that CTA  will  probe a large region of the parameter space of the IDM. The CTA sensitivity lines were taken from  \cite{Silverwood:2014yza} and \cite{Lefranc:2015pza}. They include the Galactic diffuse emission and systematics errors ($1\%$ for the upper line, $0.3\%$ for the lower line), and assume a Einasto DM profile. Notice that for this model the sensitivity of CTA largely surpasses the current and projected bounds from Fermi and HESS.} 
\label{fig:1}
\end{figure} 

In this paper we explicitly show that this is not the case. Even for a Einasto profile, the CTA will be able to significantly probe the viable parameter space  of one of the simplest and best known models of dark matter, the Inert Doublet Model (IDM) \cite{Barbieri:2006dq,Ma:2006km,LopezHonorez:2006gr}. In this model, the Standard Model is extended with a second Higgs doublet that is odd under a $Z_2$ symmetry, ensuring the stability of the dark matter particle --the lightest  neutral component of the doublet. Two features make the IDM particularly sensitivity to CTA searches. First, the dark matter particle is relatively heavy, lying above 500 GeV in the so-called high mass regime (in the low mass regime the dark matter mass is instead below $M_W$). Second, the annihilation cross section is larger than the thermal one as a result of the coannihilation effects that determine the relic density via freeze-out in the early Universe. These coannihilation effects are an  unavoidable feature of this model and imply dark matter annihilation rates up to three times larger than the thermal one. Consequently, the CTA may exclude, with conservative $1\%$ systematics,  the IDM up to dark matter masses of 800 GeV, and almost the entire parameter space if $0.3\%$ systematics are achieved, 
 as illustrated in figure \ref{fig:1}.

\section{The Inert Doublet Model}

In the Inert Doublet Model (IDM) \cite{Barbieri:2006dq,Ma:2006km,LopezHonorez:2006gr}, the SM is extended with a second Higgs doublet $H_2 =(H^+ ,(H^0+iA^0)/\sqrt{2})$, which is odd under an exact $Z_2$ symmetry (all SM field are instead even). This discrete symmetry avoids tree level flavor changing neutral currents by preventing the coupling between  $H_2$ and SM fermions, and it  guarantees the stability of the  lightest neutral scalar, which constitutes a  natural dark matter candidate. In this scenario the scalar potential is written as
\bea \label{eq:potential}
\mbox{ }\hspace{-0.5cm} 
V&=& \mu_1^2 \vert H_1\vert^2 + \mu_2^2 \vert H_2\vert^2+ \lambda_1 \vert H_1\vert^4 + \lambda_2 \vert H_2\vert^4
\cr 
&&\quad 
+ \lambda_3 \vert H_1\vert^2 \vert H_2 \vert^2 + \lambda_4 \vert H_1^\dagger H_2\vert^2 + 
{\lambda_5} Re[(H_1^\dagger H_2)^2],\;
\eea 
where $H_1$ is the SM Higgs doublet, and $\mu_{1,2}^2$ and $\lambda_{i}$ are  real parameters. After the spontaneous symmetry breaking, the spectrum contains the SM Higgs boson with $m_h=125$ GeV, and three additional (odd) scalars  with masses
\bea \label{eq:masses}
  m_{H^\pm}^2 & =& \mu_2^2 +  \lambda_3 v^2,\\
  m_{A^0}^2   &=& \mu_2^2 + \lambda_{S} v^2, \\
  m_{H^0}^2   &=& \mu_2^2 + \lambda_{L} v^2, 
\eea
 where $v \approx175$ GeV is the Higgs vev, $\lambda_{S} = \lambda_3 + \lambda_4 - \lambda_5$, and $\lambda_{L} = \lambda_3 + \lambda_4 + \lambda_5$. This model has only 5 independent parameters, which can be taken to be the three masses ($m_{H^\pm}, m_{A^0}, m_{H^0}$) and the scalar couplings $\lambda_{L}$ and $\lambda_2$.  The dark matter particle is either $H^0$ or $A^0$, both give rise to the same phenomenology. For definiteness, we assume that  $H^0$  is the lightest inert particle and hence our DM candidate, $m_{H^0}< m_{A^0},m_{H^\pm}$. Being part of a $SU(2)$ doublet,  $H^0$ has weak interactions and provides a typical  example of so-called WIMP (Weakly Interacting Massive Particle) dark matter. 

The phenomenology of the IDM is quite rich and has been extensively studied in the literature --see e.g. \cite{LopezHonorez:2006gr, Gustafsson:2007pc, Pierce:2007ut, Agrawal:2008xz, Lundstrom:2008ai, Andreas:2009hj, Hambye:2009pw, Andreas:2009hj, Arina:2009um, Dolle:2009fn, Dolle:2009ft, Honorez:2010re, LopezHonorez:2010tb,Ginzburg:2010wa,Gustafsson:2012aj, Arhrib:2012ia, Klasen:2013btp,Goudelis:2013uca}. It turns out that the relic density constraint can be satisfied in two different mass regimes. The low mass regime, $\mdm\lesssim M_W$, is currently being probed by collider and direct detection experiments and will be further tested by XENON1T and the LHC in the near future. For dark matter masses between $M_W$ and $500$ GeV, the relic density is below the current bound due to the efficient annihilation into gauge bosons. The high mass regime features $\mdm>500$ GeV and is characterized by small mass splittings between the odd particles ($\Delta m\lesssim 15~\gev$).  As a result, the relic density in this regime is always dominated by coannihilation effects. This heavy regime is currently not constrained by any experiments and is the subject of our analysis.

\section{Results and Discussions}
For our analysis, we have scanned the heavy regime of the IDM and obtained a large sample of viable points. The IDM contains only four phenomenologically relevant parameters, which can be taken to be  $m_{H^0}$, $m_{A^0}$, $m_{H^+}$, and $\lambda_L$. First of all we verified that the perturbativity constraint $|\lambda_i|<1$, which is somewhat arbitrary, requires the dark matter mass to be below $3$ TeV (larger masses would be allowed if we relaxed this condition). Then we varied $m_{H^0}$ between $525$ GeV and $3000$ GeV in steps of $25$ GeV. For each step we  generated random values of $\lambda_L$ (between $10^{-4}$ and $1$, logarithmically) and values of $m_{A^0}$ and $m_{H^+}$ not far from $m_{H^0}$. On the points thus generated, we imposed  perturbativity ($|\lambda_i|<1$), vacuum stability,  collider searches \cite{Gustafsson:2012aj,Belanger:2015kga} and electroweak precision constraints. If these constraints were satisfied, we would further require the relic density, obtained via thermal freeze-out in the early Universe, to be in agreement with the observations \cite{Ade:2013zuv} and the direct detection cross section to be compatible with the LUX bound \cite{Akerib:2013tjd} --for these calculations we used micrOMEGAs \cite{Belanger:2013oya}. A point is considered viable if it satisfies all these requirements. For each dark matter mass, we generated about 100 viable models in this way, and the total set of models for different dark matter masses constitute our full sample. Given the simplicity of the parameter space, we are confident that our sample  faithfully represents the heavy regime of the IDM. In figure \ref{fig:1}, this sample of  viable models was projected onto the plane ($\mdm$, $\sigma v$), where $\sigma v$ denotes the total annihilation cross section.  

\begin{figure}[!t]
\includegraphics[width=\columnwidth]{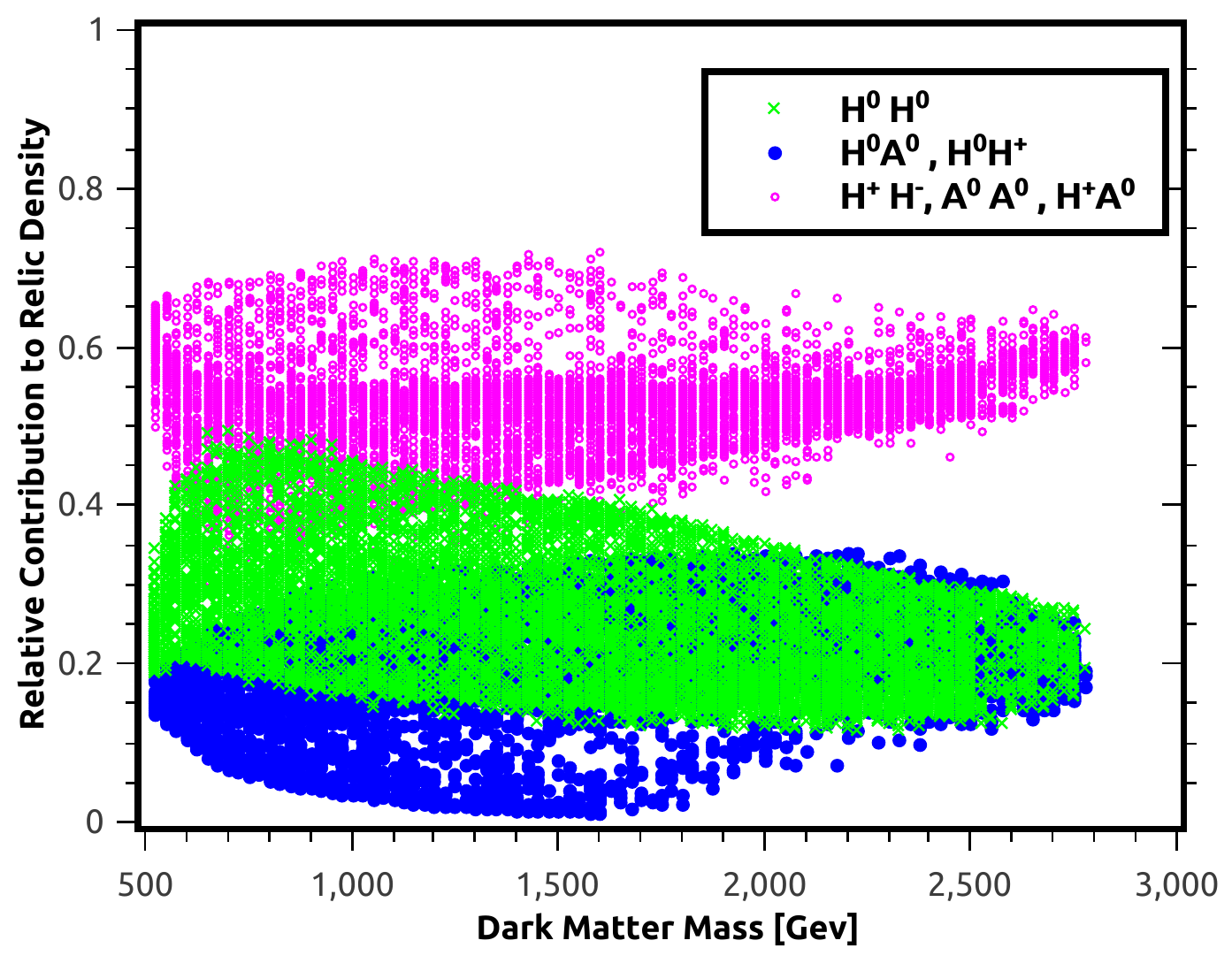}
\caption{The relative contribution to the relic density of the different annihilation and coannihilation processes. Dark matter annihilations (green points) always give a subdominant contribution to $\Omega$. Coannihilation processes involving one dark matter particle (blue points) may account for up to $35\%$ of the relic density. The dominant contribution to $\Omega$ comes from coannihilation processes not involving the dark matter particle (magenta points). All points yield the right relic abundance.} 
\label{fig:2}
\end{figure} 

In the high mass regime of the IDM, coannihilation effects play a significant role in determining the thermal relic density  in the early Universe.  A way to quantify their relevance is by considering the ratio between the coannihilation rate and  the total annihilation rate at the time of freeze-out,   $\langle \sigma v\rangle_{coann}(T_{f.o.})/\langle \sigma v\rangle_{total}(T_{f.o.})$.  This ratio defines the relative contribution of the coannihilation process to the relic density.  Figure \ref{fig:2} shows, for our sample of viable models, the relative contribution to $\Omega$ of the dark matter annihilation processes ($H^0H^0\to SM$), coannihilation processes involving one dark matter particle (e.g. $H^0H^+\to W^+Z$), and coannihilation processes not involving dark matter particles (e.g. $H^+H^-\to W^+W^-$). From the figure we see that the relic density contribution from dark matter annihilations (green points) is above $10\%$ but hardly ever reaches $50\%$ (only for $\mdm\sim 650$ GeV), and it tends to decrease with the dark matter mass. It lies below $40\%$ for $\mdm\gtrsim 1.6$ TeV and below $30\%$ for $\mdm\gtrsim 2.5$ TeV.  Dark matter annihilations, therefore, do not determine the relic density in this model --coannihilations always play an important role.  Coannihilation processes involving one dark matter particle (blue points) may account for up to $35\%$ of the relic density, lying above $10\%$ for high masses ($\mdm\gtrsim 2.5$ TeV). Coannihilation processes not including a dark matter particle  (magenta points) are the most important ones, rarely going below $40\%$ and reaching contributions above  $60\%$ over the entire range of dark matter masses. Because coannihilations play a decisive role in obtaining the observed relic density, the dark matter annihilation cross section, $\sigma v$, is \emph{not} expected to be given by the so-called thermal value ($\sigma v_\mathrm{thermal}=3\times 10^{-26} \mathrm{cm^3s^{-1}}$). Indeed, as shown in figure \ref{fig:1}, in the IDM $\sigma v$ is always larger than the thermal one and can reach values as high as $10^{-25} \mathrm{cm^3s^{-1}}$. This larger value of $\sigma v$, an unavoidable feature of the IDM, is crucial for our results as it guarantees that many viable points lie within the expected sensitivity of the CTA.   

\begin{figure}[!t]
\includegraphics[width=\columnwidth]{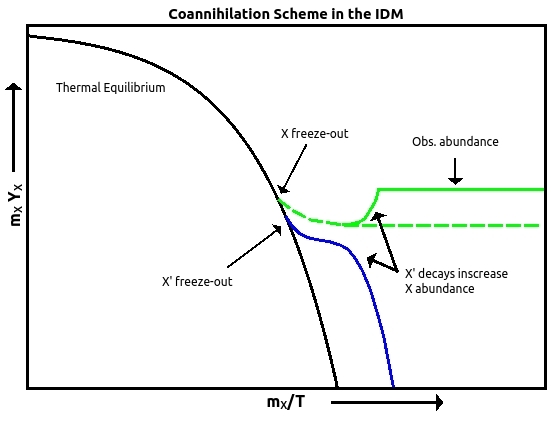}
\caption{Scheme of coannihilation effects in the IDM. Since the $H^0H^0$ annihilation cross section was large in the early universe it yielded a small dark matter abundance. Albeit, due to the presence of coannihilating particles (generically labeled as $X^{\prime}$), which share similar interactions, and thus froze-out roughly at the same time, they eventually decayed increasing the overall dark matter abundance. Since co-annihilating particles are absent today and we measure the $H^0H^0$ annihilation cross section which is large. For this reason, one can reproduce the correct relic density at freeze-out with a rather large dark matter annihilation cross section today.} 
\label{fig:scheme}
\end{figure}

It is not difficult to understand qualitatively how such a larger value of $\sigma v$ can be compatible with the observed dark matter density in the IDM. Notice that $H^0H^0$ annihilations by themselves would give rise, by virtue of the large value of $\sigma v$, to  a dark matter density below the observed value. Coannihilations, therefore, should somehow increase the dark matter density. In the IDM this process takes place in two steps. First, the coannihilating particles ($H^{\pm}, A^0$) freeze-out almost independently of $H^0$ and, since  they have practically the same interactions and the same mass, at roughly the same temperature and with a similar abundance. Second, after the freeze-out of all the particles involved --$H^0, A^0, H^\pm$-- the heavier particles decay into the lightest odd particle ($H^0$), increasing the dark matter abundance. In other words, the decays of $H^\pm$ and $A^0$ provide an extra source for  dark matter particles. This process is depicted in figure \ref{fig:scheme}. Let us emphasize, though, that this figure is only meant as an illustration  and was not obtained as a solution of the Boltzmann equation. 

\begin{figure}[!t]
\includegraphics[width=\columnwidth]{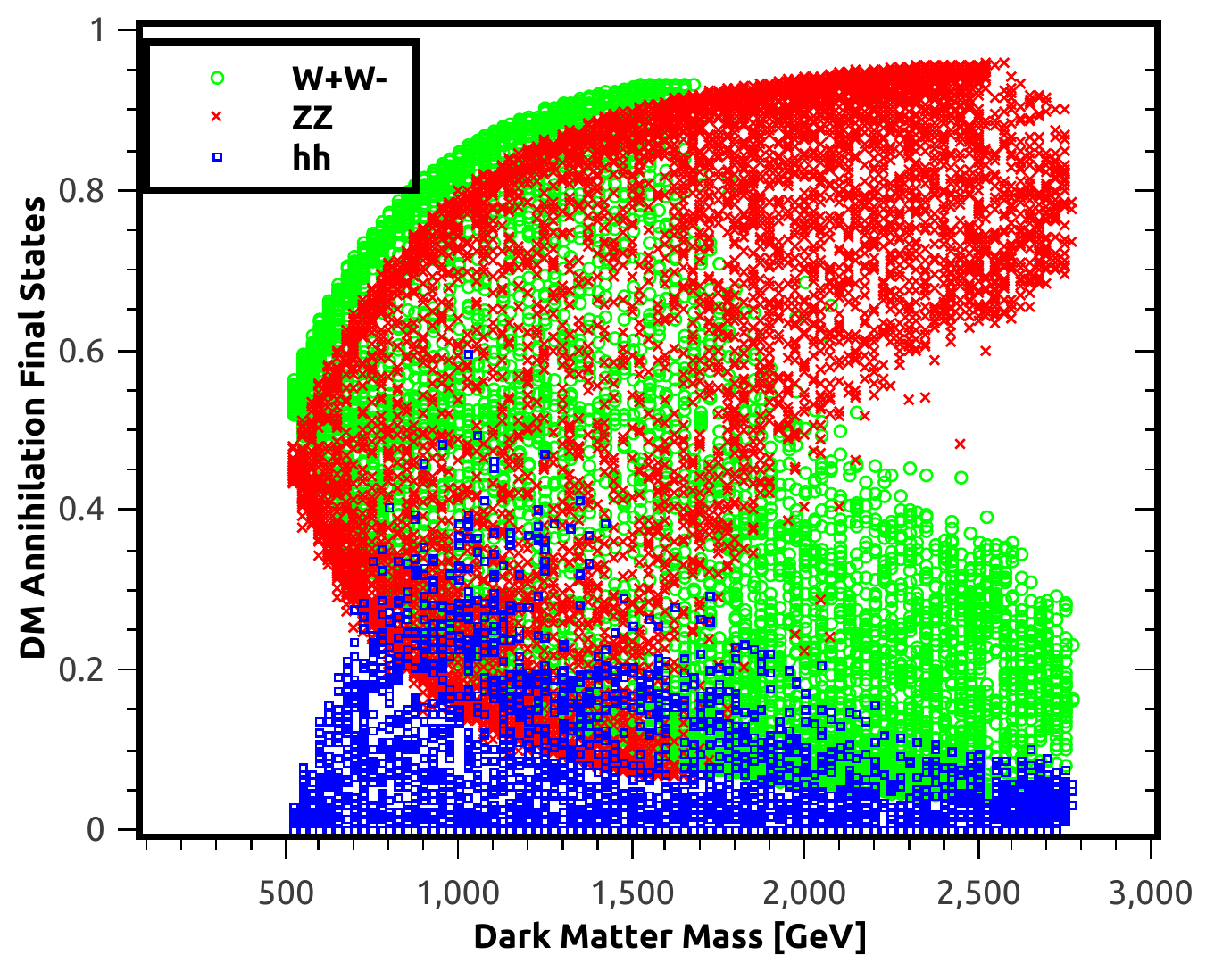}
\caption{The branching ratio of the dark matter annihilation into different final states. Only three final states are relevant in this model: $W^+W^-$ (green points), $ZZ$ (red points), and $hh$ (blue points). A tiny contribution ($<4\%$) from the $t\bar t$ final state is not shown in this figure. Notice that the annihilation cross section is dominated by gauge boson final states. } 
\label{fig:3}
\end{figure}

The gamma ray flux produced by dark matter annihilations depends not only on the mass and the annihilation cross section but also on the final states from dark matter annihilations. In the heavy regime of the IDM, dark matter particles annihilate dominantly into three different final states: $W^+W^-$, $ZZ$, and $hh$. Their relative contributions to the total annihilation cross section are displayed in figure \ref{fig:3} for our sample of viable models. The $t\bar t$ final state never accounts for more than $4\%$ of the total rate and is not shown in the figure. Notice that the contribution from Higgs final states ($hh$) hardly goes above $30\%$ \footnote{Let us stress that our limits have been obtained assuming annihilation into gauge bosons and  would not strictly apply to  the few models with a significant annihilation into $hh$. Albeit, the latter yields a harder gamma-ray spectrum and therefore our bounds simply rather conservative in this case.}. In consequence, dark matter annihilations in the IDM are largely dominated by gauge boson final states. Regarding the relation between the $W^+W^-$ and $ZZ$ branching fractions, either can dominate for masses below $2$ TeV or so. For larger masses, it is the $ZZ$ final state that gives the dominant contribution. In any case, the gamma ray yield produced by dark matter annihilations into $W^+W^-$ and $ZZ$ are practically indistinguishable, so it does not matter if it is one or the other that dominates. The relevant point is that dark matter annihilates mostly into gauge boson final states. That is why in figure \ref{fig:1} we have compared the viable parameter space of the IDM against the expected CTA sensitivity for dark matter annihilation into $W^+W^-$. 

Let us now briefly review the assumptions that were used to obtain the CTA sensitivity regions shown in figure \ref{fig:1} --see \cite{Silverwood:2014yza,Lefranc:2015pza} for further details. First of all, they rely on  the CTA configuration known as Array I, which consists of 3 large, 18 medium, and 56 small telescopes. The limits themselves are quite realistic as they take into account not only the expected backgrounds from cosmic rays and the galactic diffuse emission (GDE), but also systematic errors at the   $1\%$ (upper line) or $0.3\%$ (lower line) level. They further assume  100 (upper line) or 500 (lower line) hours of observations of the Galactic Center, and a dark matter distribution given by the Einasto profile. In addition, they were derived using a morphological analysis (rather than the Ring method) that exploits the shape differences between the GDE and the dark matter annihilation signal. As shown in figure \ref{fig:1}, the expected CTA sensitivity from \cite{Silverwood:2014yza} (upper line) cuts through the viable  parameter space of the IDM. In particular,  all the viable models with  $\mdm\lesssim 800~\gev$ lie above the CTA sensitivity line. The CTA may, therefore,  entirely exclude that region of the parameter space. At higher masses, the CTA can probe  a significant fraction of the remaining  parameter space over the entire range of dark matter masses. 

The lower CTA sensitivity line, obtained from \cite{Lefranc:2015pza}, assumes instead $0.3\%$ systematics and was derived under slightly different assumptions for the cosmic ray background and the galactic diffuse emission. As can be seen in figure \ref{fig:1}, essentially the entire viable parameter space of the IDM is within the expected CTA sensitivity region in this case. To be cautious, in our presentation we have mostly relied on the weaker limit from \cite{Silverwood:2014yza}. In this way we ensure that our main result is robust: the CTA will genuinely challenge the IDM.

In figure \ref{fig:1} we also display the current and projected sensitivity of Fermi-LAT, and the projected sensitivity of HESS II. Current bounds by Fermi-LAT, which are based on 6 years of observation of dwarf galaxies \cite{Ackermann:2015zua}, do not yet reach the viable region. And after 10 years, Fermi-LAT will be able to probe only a small fraction of the viable parameter space --at low masses and high cross sections.  Similarly, HESS II, with 250 hours of observation of the galactic center \cite{Lefranc:2015vza}, will be able to test only the region with very high cross sections around $\mdm\sim 1$ TeV.  As seen in the figure, both of these regions are well within the expected CTA sensitivity. Thus, the CTA will provide the most stringent constraints on this model.

In our analysis we did not include two effects that may strengthen our results: gamma ray spectral features \cite{Bringmann:2007nk} and the Sommerfeld enhancement \cite{Hisano:2006nn}. For the IDM, the relevance of a gamma ray spectral feature from the annihilation of dark matter into $W^+W^-\gamma$ was studied in \cite{Garcia-Cely:2013zga,GarciaCely:2014jha}. They  pointed out that such a feature is generically expected in the heavy regime of this model. Another feature comes from the direct annihilation of dark matter into two-photons at one loop, which produces a gamma ray line at $E_\gamma\sim \mdm$ --see \cite{Gustafsson:2007pc} for a related work of this effect in the low mass regime of this model. As a result,  the gamma ray spectrum receives  additional contributions at energies close to the DM mass, where the CTA is  more sensitive.  Determining the CTA sensitivity region including this new contribution to the gamma ray spectrum is, however, non-trivial and lies beyond the scope of the present paper. It is clear, though, that the sensitivity can only increase (the CTA line will move to lower values of $\sigma v$), so that the region that can be probed by the CTA may actually be larger than shown in figure \ref{fig:1}. The Sommerfeld enhancement, on the other hand, may increase the present value of the annihilation cross section, moving the viable points upwards in figure \ref{fig:1} without modifying the CTA sensitivity line. The final result would again be that the constraints can only be stronger than shown  in our figure. In any case, the Sommerfeld enhancement is not expected to be that large in this model \cite{Cirelli:2007xd} and a detailed analysis  is currently in progress \cite{talkSEIDM}. Summarizing, the results shown in figure \ref{fig:1} can actually be considered as conservative, as they do not include the Sommerfeld enhancement nor gamma ray spectral features, both of which are expected to improve the CTA constraints on the IDM.  

\subsubsection*{Addendum}
After this paper was submitted to the arxiv, the work in progress referred to in \cite{talkSEIDM} appeared as \cite{Garcia-Cely:2015khw}. In it, the authors did a more detailed analysis of the detection prospects of the IDM in  Cherenkov Telescopes, including, in particular, the Sommerfeld enhancement and the gamma ray features. Our results pretty much agree with theirs, which  state that ``...a significant part of the viable models of our scan can be potentially probed by CTA".  

\section{Conclusions}
We demonstrated that the CTA can realistically probe the viable parameter space of the Inert Doublet Model, one of the simplest and best known models of dark matter. Specifically, the CTA may exclude dark matter masses up to 800 GeV and constrain a large fraction of the models with heavier masses. This strong bound is the result of two unique properties of the IDM: its relatively heavy dark matter particle ($\mdm>500~\gev$), and a dark matter annihilation rate  that is \emph{always} larger than the thermal one. Such a higher annihilation rate,  a generic and unavoidable feature of this model, is a consequence of the coannihilation effects that determine the relic density via thermal freeze-out in the early Universe.  To obtain the above mentioned results, we first scanned the  parameter space of the IDM (in the heavy regime) and obtained a large sample of models consistent with all known theoretical and experimental bounds. We then compared these models against the expected sensitivity of the CTA to a dark matter annihilation signal from 100 hours of Galactic Center observation assuming a morphological analysis that includes both systematics errors and galactic diffuse emission. We also showed that the CTA expected limit widely surpasses current and projected constraints from Fermi-LAT and HESS on this model. 

\section*{Acknowledgement}
The authors thank Christoph Weniger, Paolo Panci, Camilo Garcia and Aion Viana for correspondence. CY is supported by the Max Planck Society in the project MANITOP.

\bibliography{darkmatter}

\end{document}